# A Consistent Microlensing Model for the Galactic Bar

HongSheng Zhao
Max-Planck-Institute für Astrophysik, 85740 Garching, Germany
Email: hsz@MPA-Garching.MPG.DE

R. Michael Rich
Dept. of Astronomy, Columbia University, NY, NY 10027, USA
Email: rmr@cuphyd.phys.columbia.edu

David N. Spergel
Princeton University Observatory, Princeton, NJ 08544, USA
Email: dns@astro.princeton.edu

## ABSTRACT

We compute a microlensing map for the Galactic bar. The predicted event rate and event duration distribution are consistent with the 55 events recently reported by the MACHO and OGLE collaborations. Most of the events are due to lensing by stars in the near end of the bar. Lens mass functions with about 30-60% of lens mass as brown dwarfs are rejected at $2 - 6\sigma$ levels. To make our model useful for other workers, we tabulate the bar's optical depth and average event duration (scaled to $1 M_\odot$ lenses) on a grid of Galactic coordinates. The distance and the proper motions of the lens and the source are derived from a consistent dynamical model of the stellar bar, which has originally been built to fit data on the stellar light and stellar/gas kinematics of the bar. We explore several alternative models and we find that our standard model best matches observations.

*Subject headings*: Galaxy: structure - Galaxy: Kinematics and Dynamics - gravitational lensing - stars: low-mass, brown dwarfs - dark matter

## 1. Introduction

Recent microlensing observations towards the Galactic bulge imply a larger optical depth than can be accounted for by lensing by an axisymmetric bulge and disk (Udalski et al. 1994, Alcock et al. 1995). Paczyński et al. (1994) proposed that this excessive optical depth was due to lenses in the near end of the Galactic bar, which has been discovered in a number of observations (see Gerhard 1995 for a review). Although the idea is very attractive, they lacked a good bar model to compare with the observation.

To calculate the microlensing event rate due to the bar requires the distance and proper motion distributions of stars in the Galactic bar. Unfortunately most of our data in the bar probe rather different projections of the phase space, which include the surface light map from COBE (Weiland et al. 1995) and many radial velocities of bulge stars (see Zhao 1995). But one can still simulate the unknown dimensions of the phase space effectively by "observing" a stellar dynamical model which both fits the available observations with a steady state distribution function.

As part of his doctoral thesis, Zhao (1994) built a dynamical model for the Galactic bar with a variant of Schwarzschild's (1979, 1982) method to study kinematic data of the bar. In this model, stellar orbits in a fixed potential are populated to be consistent with the COBE map and kinematics at Baade's window. Later on Zhao, Spergel & Rich (1995, hereafter ZSR) built a microlensing model based on this dynamical model, and confirmed Paczyński et al.'s (1994) suggestion that the excess lensing events could be due to the bar. They were also able to place a preliminary constraint on the mass function of the lenses in the bar: the typical 20 days Einstein radius crossing time scale of the 9 events from the first two years of the OGLE observation already favors a mass function dominated by faint main sequence stars rather than by brown dwarfs.

This paper further extends ZSR's model, and presents predictions of lensing rate in many (observed or plausible) Galactic bulge fields. It improves over our earlier work in several ways: First, the microlensing model is based upon an improved stellar orbit model (Zhao 1995), which fits the boxyness of the bulge seen in the COBE image better than the previous model (Zhao 1994). The new model is also consistent with the potential of the bar and the disk, and has been checked for stability. Second, our analysis is based on the enlarged microlensing data set rather than the earlier 9 OGLE events. We fit event duration distribution from the combined MACHO and OGLE events (55 events from 37 lines of sight) with their respective efficiencies (Bennett et al. 1995, Udalski et al. 1995). These improvements result in much stronger constraints on the lower end of the bulge mass function.

## 2. The Bar Model

The microlensing rate depends on the underlying mass model. We call the stellar model by Zhao (1995) our "standard" model, which is made to reproduce a modified G2 model. The G2 model is one of the three-dimensional luminosity models that Dwek et al. (1995) fit to the COBE map. We have found that the rate is most sensitive to the angle between the long axis of the bar and our line of sight and the total mass of the bar. If one moves the bar's long axis closer to our line of sight while holding the projected light distribution and the velocity distribution of the bar fixed, one effectively increases the projected mass density and the typical source-lens separation, which lead to a larger optical depth and a longer event duration. Increasing the bar mass alone will also lead to a larger optical depth. These dependences can be seen by comparing several of the different 3D luminosity models fit by Dwek et al. (1995) to the COBE light distribution. Their two best-fit models, the G2 model with a bar angle of $13^o - 20^o$ and the E3 model with an angle of $40^o$, have very similar total luminosity and mass-to-light ratio. However, they make very different predictions for microlensing; for the same fixed stellar mass function, the E3 model predicts 50% smaller optical depth than the G2 model and a 70% shorter event duration.

Fortunately, the two models also make two different predictions for gas kinematics. It is informative to compare the morphology of the closed $x_1$ orbits of the bars with the profiles of the HI and CO maps of the Galaxy. Binney et al. (1991) argue that if gas clouds move on non-self-intersecting closed orbits as a result of the pressure force and dissipational collisions, the non-self-intersecting $x_1$ orbits would trace the profile of the HI map. On the other hand CO may form only in the shocked region near where closed $x_1$ orbits develop self-intersecting loops. Binney et al. show that simulated $l - v$ diagrams of $x_1$ orbits are very sensitive to global parameters of the potential, in particular, the angle of the bar, the pattern speed and the mass of the bar. While the analysis of Binney et al. is based on simple plausible potentials, our aim is to see if the bar potentials which have been constrained by the *COBE* map and the stellar velocity data are also consistent with the gas kinematics.

We compute the closed $x_1$ orbits in the potentials of the G2 and E3 models; both potentials include a disk and a nucleus as in the stellar model. The modified *G2* model has a bar angle of $20^o$ and a bar mass $M_{bar} = 2.24 \times 10^{10} M_\odot$. The modified E3 model has a bar angle of $40^o$ and a bar mass $M_{bar} = 2.1 \times 10^{10} M_\odot$. Both bar models have a pattern rotation speed $60\xi^{1/2}$ km/s/kpc, where $\xi = M_{bar}/(2 \times 10^{10} M_\odot)$. These values of the bar parameters have been constrained to reach consistency with their potentials using the Schwarzschild method; the consistency for the E3 model is somewhat worse than for the G2 model.

Figures 1 overplots the simulated $l - v$ diagrams of the bar models on the HI map, which is based on a data cube in FITS format kindly made available by Harvey Liszt. As in the Binney et al. model, the non-intersecting $x_1$ orbits in the G2 model well match the rise of the terminal velocity curve in the HI map from $l = 0^o$ to $l = 2^o$ and the fall from $l = 2^o$ to $l = 10^o$ Figure 1). The nearly cusped $x_1$ orbit (the solid line loop in Figure 1) matches the vertical edge of the parallelogram at $l = 2^o$ in the CO map (see the reproduction of Bally et al. 1987's data in Figure 2 of Binney et al.). Since a good match to the vertical edge requires that one of the sides of the cusp orbit nearly coincide with the $l = 2^o$ line-of-sight, this fixes the angle of the bar near $16^o$ and the corotation radius to $\approx$ 3kpc (Binney et al. 1991). The amplitude of the terminal velocity curve, which goes roughly as $M_{bar}^{0.5}$, also constrains the mass of the bar. Surely enough, a bar with a large angle (the E3 model) appears to be poorer in fitting the profile of the HI map, and its parallelogram does not have a vertical edge (the lower panel of Figure 1). Also the edge of the parallelogram are pushed outward to $|l| > 2^o$ for models with the same potential and perspective as the G2 model but a much slower pattern rotation. In short, our standard model fits the HI and CO map of the Galaxy. And we confirm Binney et al. (1991) conclusion that the bar angle is near $16^o$. This bar angle together with the requirement of triaxial reflection symmetry also fixes the density profile of the bar (Binney and Gerhard 1995).

Figures 5 and 6 in Zhao (1995) also show that the orbit distribution in the standard model is consistent with the input density model of the bar and consistent with the COBE map and stellar kinematics. Taking into account that it is also a smooth and stable model, we argue that our standard model gives so far a most consistent description of the Galactic bar, and that it is a reliable model to interpret the microlensing data.

## 3. Predictions of Microlensing Rates

We convert the stellar model to a microlensing model. We first construct an N-body realization of the bar with about $10^6$ stars distributed on a few hundred orbits, with random phase; the number of stars on each orbit is proportional to the weight of the orbit. A double-exponential disk with a central hole of 3 kpc radius is also included (Paczyński et al. 1994) and the velocities of the disk stars are assigned with a narrow 20 km/s isotropic Gaussian distribution on top of the flat rotation curve. As the bar ends within its corotation 3 kpc, the bar stars and the disk stars do not overlap, which allows one to identify a star as a disk star or a bar star just according to its distance without ambiguity. We then "observe" this N-body system to simulate the transverse velocity distribution of lenses and sources. The distance distributions of lenses and sources are prescribed by the G2 volume density model without a nucleus for the bar and the hollow double-exponential disk model.

Note that we do not take the density of bar stars from the N-body system directly, which is known to be noisier, slightly shorter than the Dwek et al. model (see Figure 4 in Zhao 1995) and more concentrated to the center due to the small nucleus which has 5% of the bar's mass. These may cause us to overestimate the bar's optical depth by $5-10\%$ at regions outside $2-3^o$ of the center. However, most of the uncertainty in optical depth is expected to come from the inner few kpc of the disk. Based on this bar and disk model, we make a microlensing map in 20 by 20 degrees sky region around the Galactic center.

Predictions of the microlensing event distribution are a function not only of the stellar phase distribution but also of the lens distribution in the disk, the mass spectrum of the lenses and the detection efficiency. Several of our colleagues, including Bohdan Paczyński, Kim Griest, David Tytler, encouraged us to present the lensing rates for our standard model in the special case where all of the lenses are in the bar, their masses are all $1M_\odot$ and the detection efficiency is 100%. The event duration distribution of this single mass model (SM model) depends only on the bar stellar phase space distribution. We make the SM model for the bar available electronicly in tabular form so that one can use them to make predictions after simply convolving a mass function and the detection efficiency and adding a disk.

Some microlensing calculations, such as estimating the detection rate of planetary systems or binary systems require a model for the distribution of the relative velocity between a lens and a source as function of their positions. These calculations require more information than the event duration distribution. We find that in the case that both the lens and the source are in the bar the relative proper motion has a distribution that of a 2D Boltzmann distribution, that is, $f(v)dv = \frac{v}{2\mu^2} \exp(-\frac{v^2}{4\mu^2})dv$, where $v$ is the source-lens relative proper motion speed projected to a distance of 8 kpc. Typically $\mu$, which is certain averaged proper motion dispersion of the lens and the source, is of order 100 km/s. A more meaningful and convenient number to characterize the velocity distribution is the mean relative speed $<v> \equiv \int_0^\infty v f(v) dv = \sqrt{\pi}\mu$. This mean relative speed also enters into the equation (11) of Kiraga and Paczyński (1994) for the total event rate. It is a function of four variables, namely, the lens and source distances and the sky direction. Its value on a grid of lens and source distances at Baade's Window is given in Table 1, together with the lens or source mass density on the grid. Values at other sky directions are available via ftp (see Conclusions).

Constraints on the mass spectrum of lenses require only the SM model time scale distribution, where distance and velocity informations have already been convolved together. Even in this reduced form, the normalized event time scale distribution curve for each line-of-sight, the model contains too much information to deliver easily in tabular form.

We decide to further compress the information by fitting each event duration distribution curve with a simple functional parametrization, and tabulate the obtained parameters on a grid of the Galactic coordinates $(l, b)$. The key parameters are the optical depth $\tilde{\tau}$ (due to the bar lenses) and the median event duration, $\tilde{T}$. Both $\tilde{\tau}$ and $\tilde{T}$ are functions of $(l, b)$. For all lines of sight we use the following parametrization, which fit the lensing probability distribution reasonably well:

$$\tau(\leq t_0) = \tilde{\tau}(1 + C\tilde{T}^2 t_0^{-2})^{-n}, \tag{1}$$

where we set the additional parameters $n = 1/\epsilon$, $C = 2^\epsilon - 1$ and $\epsilon = 0.1 + 0.05\tilde{\tau}$. The function $\tau(\leq t_0)$ denotes the cumulative optical depth contributed by events with Einstein radius crossing time shorter than $t_0$. So $\tau(\leq 0) = 0$, $\tau(\leq \infty) = \tilde{\tau}$ and $\tau(\leq \tilde{T}) = 0.5\tilde{\tau}$. It is also simply related to the differential event duration distribution $P(t_0) \equiv \frac{d\Gamma}{d\log t_0}$ for a single mass lens by $P(t_0) = \frac{2\ln 10 \, d\tau(\leq t_0)}{\pi dt_0}$. Figure 2 compares the parametrized distribution (solid lines) with directly computed event differential distribution (the symbols) at three lines of sight of the bulge. The parametrization is valid to within 20%.

Both $\tilde{\tau}$ and $\tilde{T}$ are tabulated in Table 2. The bar's optical depth $\tilde{\tau}$ is approximately an exponentially decreasing function of the distance to the Galactic Center, similar to the dependence of the bulge surface brightness. The median event duration, $\tilde{T}$, for solar-mass lenses is between 30 days and 45 days, increasing gradually with the distance from the Galactic Center. The time scales for 5% and 95% of the optical depth are at about $0.4\tilde{T}$ and $4\tilde{T}$, insensitive to the value of $\epsilon$, which takes the value between 0.1 to 0.4. Note $\tilde{\tau}$ also differs systematicly at 10% level between negative longitudes and positive longitudes. However, detecting this 10% effect will require a larger data set than is currently available. The predicted variation in $\tilde{T}$ is even smaller: less than 5%.

## 4. Constraining the Stellar Mass Function

In this section, we compare the predictions of our standard bar model to the MACHO and OGLE observations (Bennett et al. 1995, Udalski et al. 1994) for several different mass functions.

The two collaborations have observed a total of 55 *independent* events without double counting OGLE #1 and the famous binary event OGLE #7. The distribution of event durations seen by the two experiments is shown as a histogram in Figure 3.

We can reject models with 30 − 60% of the mass in the form of brown dwarfs at the 2 − 6$\sigma$ level. Since the typical lens mass in these models is about $0.1 M_\odot$, predicted events are too short and somewhat too many. Although this result is insensitive to the functional form of the mass function, Figure 3 demonstrates it by a Salpeter mass function $dN/dm \propto m^{-2.25}$ with the lower cut-off at 0.04, 0.03, 0.02 or 0.01 solar masses. These cutoffs correspond to 27%, 35%, 44% or 57% brown dwarfs, which are rejected at $0.06(2\sigma)$, $0.005(3\sigma)$, $7 \times 10^{-5}(4.5\sigma)$ or $1 \times 10^{-8}(6\sigma)$ confidence level by a K-S test with the data.

We can also reject some mass functions with mostly bright lenses. Gould, Bahcall & Flynn (1995) fit a stellar mass function to their recent Hubble observations, which includes stars from $0.1 M_\odot$ to $2 M_\odot$ with the average lens mass at $0.5 M_\odot$ (Han and Gould 1995). This mass function produces events which are too long and is rejected by a K-S test at the $2 \times 10^{-4}(4\sigma)$ confidence level. It also produces significantly too few events (only 37 events). Lowering the upper mass cutoff from $2 M_\odot$ to $1 M_\odot$ still gives a very small K-S probability $5 \times 10^{-3}(3\sigma)$. As a comparison, a Salpeter mass function between $0.1 M_\odot$ and $10 M_\odot$ would produce similarly too few events (40 events), but would not be rejected by the K-S test. It is perhaps not very surprising that the typically solar mass main sequence stars (between $0.6 M_\odot$ and $1.4 M_\odot$) seen in the local disk and the bulge do not appear as lenses. As Kamionkowski (1995) notes, events with massive main sequence stars as foreground lenses and main sequence stars as background sources would have non-standard lensing light curves and would not be included as lensing events by either MACHO or OGLE, which were mainly targeting dark or very faint lenses.

The data are largely consistent with a range of models with a small amount of brown dwarfs (less than 20%) or with the number averaged lens mass somewhat above the brown dwarf limit (between $0.15 M_\odot$ to $0.3 M_\odot$). Although the microlensing predictions are insensitive to the exact lower and upper cutoffs and functional form of the mass function, it appears that the mass function needs to be either flat or truncated near the brown dwarf limit. This is shown by the good match to the data by both the solid line and the dashed line in Figure 3.

The dashed line is for the mass function from Kroupa, Tout & Gilmore (1990) which has a functional form of Miller and Scalo (1979). It is Gaussian above $0.35 M_\odot$ and flat below $0.35 M_\odot$ based on fitting their sample of local disk stars. We also applied a $0.01 M_\odot$ lower cutoff and a $0.6 M_\odot$ upper cuffoff. The mean lens mass is $0.15 M_\odot$. As their model always prescribes very few (less than 15%) brown dwarfs, it can successfully reproduce the event distribution independent of the details of the lower and upper mass cutoffs.

The solid line is for a Salpeter mass function with the lens mass between $0.08 M_\odot$ and $0.6 M_\odot$, which fits the data very well. This is our preferred mass function in estimating microlensing rates. It predicts 40 events due to lenses in the bar with a mean duration of 16 days, and 14 events due to lenses in the disk with a longer mean duration of 30 days.

Together they predict 45 events for the MACHO fields and 9 events for the OGLE fields. The total distribution accounts for both the peak of the event duration distribution and the total number of events for the MACHO and OGLE data set together. The upper panel also shows that it is also consistent with the OGLE event distribution. The average lens mass in this model is $0.18 M_\odot$.

Our standard model predicts that the event duration should be larger in fields further away from the Galactic center. The optical depth of the bar decreases steeply from the center, so the contribution to the optical depth from even the truncated disk starts to exceed the former at $8°$ from the center. Since the duration of a bar event is longer further away from the center (see Table 2), and the duration of a disk lensing event is typically longer than that of a bar event, the model predicts an increase of event duration away from the center of about 0.02 dex in $\log t_E$ per degree, or about 1 day per degree. The current sample is not yet large enough to definitively detect this effect.

Our standard model also predicts that the lensing optical depth of the Galaxy should decrease outward closely following the drop of the surface density. And when source stars are red clump stars, we expect a larger optical depth and a longer duration. This is because microlensing of clump stars at the far end of the bar and/or involving a bright massive main-sequence (disk or bar) lenses can still be registered while the same is not true for the numerous less bright bulge main sequence source stars. These effects may have already been seen: Bennett et al. (1995) found the optical depth over all of their observed fields is $3.0^{+1.5}_{-0.9} \times 10^{-6}$ when the sources are bulge red clump giants and $1.58^{+0.35}_{-0.28} \times 10^{-6}$ for the full sample. Comparing the optical depth at MACHO fields $|b| < 3.5°$ with at $|b| > 3.5°$, they also found a drop of the optical depth with the latitude for the bulge red clump sources from $5.2^{+3}_{-1.9} \times 10^{-6}$ to $1.2^{+0.5}_{-0.6} \times 10^{-6}$.

If we assume the distribution and efficiency of the red clump stars are identical to the general stellar population, our standard model predicts an optical depth of $2.1 \times 10^{-6}$ for $|b| < 3.5°$, $1.3 \times 10^{-6}$ for $|b| > 3.5°$ and $1.5 \times 10^{-6}$ for all the MACHO fields respectively. Assuming that the red clump stars can be seen throughout the bar and using the corresponding MACHO efficiency, we predict between 15 to 20 events for the red clump stars depending on whether the disk is full or truncated. The optical depth for the clump stars is $2.2 \times 10^{-6}$ for the truncated disk model and $2.8 \times 10^{-6}$ for the full disk model. The numbers are in rough agreement with the observed 13 events for the clump source stars, but are systematically smaller.

As noted by Han and Gould (1995) and Mao (1995), the 13 clump events include about 3 very long events (longer than 70 days), which is significantly more than predicted by Han and Gould's model and our standard model. We estimate that about half of the optical depth of the clump stars are due to these long events. As these events also seem to be at fields relatively closer to the Galactic plane (see Figure 1 of Bennett et al. 1995), one suspects that they are associated with a massive disk lens and a small source-lens relative proper motion. However, even when we include 10 solar mass lenses, remove the disk truncation, consider both disk and bulge sources, and increase the local density and decrease the radial scale length of the disk within plausible range, our model predicts about 1 event longer than 70 days for the red clump sample. This discrepancy also translates to the spatial distribution of optical depth of the clump stars, as the long events generally contribute more to the optical depth than the short events, and the long events are closer to the plane. In short, although the model is in overall agreement with MACHO and OGLE observations, and can qualitative explain the bias of the red clump stars, a more detailed study of the long events and their relation with clump stars remains to be done.

## 5. Conclusions

Several bar models are built for the Galaxy and compared extensively with observations of the stellar light, stellar/gas kinematics as well as the microlensing data. Only our standard model, namely the modified G2 model, is favored by all these independent constraints. We apply this model to the now 55 microlensing events towards the Galactic bulge from the MACHO and OGLE data set together.

Our standard model is consistent with the MACHO and OGLE observations if less than 20% of the mass are in brown dwarfs. This argues for a truncation or turn-over of the mass function at the lower end, which is consistent with the mass function of local disk stars (Kroupa et al. 1990) and the recent detection of only two brown dwarf candidates in the Pleiades young open cluster (Rebolo et al. 1995, Basri et al. 1995). Mass functions dominated by brown dwarfs (60% in mass) would produce many more very short time scale events than are observed, and are strongly ruled out (at $6\sigma$ level).

The model does not explain the excess of long duration events from MACHO, and their relation with the red clump stars.

In our model, most of the lenses are in the front end of the Galactic bar, which points nearly towards us. The model predicts that the lens duration will increase away from the Galactic center and that the lens optical depth will roughly trace the observed surface brightness profile and decline rapidly away from the center. The predicted microlensing map and the tabulated results for our single mass model are available electronically at anonymous ftp://ibm-1.MPA-Garching.MPG.DE/pub/hsz/.

HSZ acknowledges suggestions on the presentation of the model from Kim Griest, Bohdan Pazcyński and David Tytler. HSZ also thanks Simon White and Shude Mao for their comments on an earlier version of the manuscript. RMR acknowledges support from NASA grant NAGW-2479. DNS is partially supported by NSF grants AST91-17388, NASA grant ADP NAG5-2693 and by the GC3 collaboration.

## REFERENCES

Alcock, C. et al. 1995, ApJ, 445, 133

Bahcall, J.N., Flynn, C. & Gould, A. 1992 ApJ, 389, 234

Bally, J., Stark, A. A., Wilson, R. W., & Henkel, C. 1987, ApJS, 65, 13

Basri, G. et al. 1995, Bull. Am. Astro. Soc. (conf. abstr.) in press

Becklin, E.E. & Neugebauer, G., 1968, ApJ, 151, 145.

Bennett, D.P., et al. 1995, preprint MACHO-Bulge-002; "Microlensing Workshop" Jan. 13-15, 1995, Livermore.

Binney, J.J., & Gerhard, O.E., 1995, preprint SISSA astro-ph/9508115, submitted to MNRAS

Binney, J.J., Gerhard, O.E., Stark, A.A., Bally, J., Uchida, K.I. 1991 MNRAS, 252, 210

Dwek, E. et al. 1995, ApJ 445, 716

Gerhard, O. 1995, in IAU symposium 165, "Unsolved Problems in the Milky Way", ed. L. Blitz (Dordrecht:Kluwer)

Han, C. and Gould, A. 1995, preprint OSU-TA-5/95

Kamionkowski, M. 1995, ApJ, 442, L9

Kiraga, M. & Paczyński, B. 1994, ApJ, 430, L101

Kroupa, P., Tout, C. and Gilmore, G. 1990, MNRAS, 244, 76

Mao, S., 1995, private communications

Miller, G. and Scalo, J.M. 1979, ApJS, 41, 513

Paczyński, B., Stanek, K.Z., Udalski, A., Szymański, M., Kaluźny, J., Kubiak, M., Mateo, M. & Krzemiński, W. 1994, ApJ, 435, L63

Paczyński, et al. 1994, AJ 107, 2060

Schwarzschild, M. 1979, ApJ, 232, 236

Sharples, R., Walker,A., & Cropper, M. 1990, MNRAS, 246, 54

Spaenhauer, A., Jones, B.F. & Whitford, A.E. 1992, AJ, 103, 297

Rebolo, R. et al. 1995, Nature, 377, 129

Udalski, A., Szymański, M., Stanek, K.Z., Kaluźny, Kubiak, M., Mateo, M., Krzemiński, W., Paczyński, B. & Venkat, R. 1994, Acta Astronomica, 44, 165

Weiland, J. et al. 1994, ApJ, 425, L81

Zhao, H.S. 1994, Ph.D. thesis, Columbia University, New York

Zhao, H.S., Spergel, D.N., & Rich, R.M. 1994, AJ, 108, 6

Zhao, H.S., Spergel, D.N., & Rich, R.M. 1995, ApJ, 440, L13,

Table 1. Predicted mean relative velocity $<v>$ between lens and source at Baade's Window

| $D_s$ | $\rho$ | $D_l = 6.00$ | 6.25 | 6.50 | 6.75 | 7.00 | 7.25 | 7.50 | 7.75 | 8.00 |
|---|---|---|---|---|---|---|---|---|---|---|
| 6.00  | .65  | ... | ... | ... | ... | ... | ... | ... | ... | ... |
| 6.25  | .86  | 181 | ... | ... | ... | ... | ... | ... | ... | ... |
| 6.50  | 1.06 | 177 | 179 | ... | ... | ... | ... | ... | ... | ... |
| 6.75  | 1.19 | 187 | 191 | 188 | ... | ... | ... | ... | ... | ... |
| 7.00  | 1.25 | 191 | 194 | 191 | 201 | ... | ... | ... | ... | ... |
| 7.25  | 1.24 | 199 | 202 | 200 | 209 | 207 | ... | ... | ... | ... |
| 7.50  | 1.19 | 198 | 201 | 201 | 208 | 204 | 202 | ... | ... | ... |
| 7.75  | 1.12 | 222 | 225 | 228 | 230 | 225 | 219 | 206 | ... | ... |
| 8.00  | 1.06 | 220 | 224 | 227 | 228 | 222 | 215 | 202 | 208 | ... |
| 8.25  | .99  | 216 | 219 | 224 | 226 | 218 | 210 | 193 | 200 | 193 |
| 8.50  | .92  | 228 | 232 | 240 | 238 | 228 | 216 | 196 | 198 | 189 |
| 8.75  | .84  | 235 | 238 | 246 | 244 | 233 | 219 | 198 | 197 | 188 |
| 9.00  | .74  | 251 | 253 | 262 | 259 | 245 | 230 | 206 | 199 | 189 |
| 9.25  | .62  | 248 | 252 | 262 | 257 | 243 | 227 | 202 | 195 | 184 |
| 9.50  | .50  | 226 | 229 | 238 | 234 | 222 | 207 | 181 | 180 | 169 |
| 9.75  | .38  | 231 | 230 | 239 | 238 | 223 | 208 | 181 | 179 | 168 |
| 10.00 | .27  | 217 | 214 | 220 | 223 | 208 | 195 | 168 | 171 | 160 |

Note. — For a given source distance $D_s$ (in kpc) the table gives the local density $\rho(D_s)$ (in $M_\odot pc^{-3}$) and the mean relative velocity $<v>$ (in km/s) at a set of lens distances $D_l$ (in kpc). Predictions are for bar lenses between 6 kpc and 8 kpc and bar sources between 6 kpc and 10 kpc in Baade's Window.

Table 2. Predicted spatial distributions of the optical depth $10^6\tilde\tau$ and the half-time $\tilde T$ (days) for $1M_\odot$ lenses in the bar.

| $10^6\tilde\tau, \tilde T$ | $l = \pm 1^o$ | $l = \pm 3^o$ | $l = \pm 5^o$ | $l = \pm 7^o$ | $l = \pm 9^o$ |
|---|---|---|---|---|---|
| $b = 0^o$ | 2.08, 30 | 1.72, 31 | 1.26, 31 | .87, 34 | .69, 40 |
|  | 2.21, 30 | 1.94, 30 | 1.37, 31 | .72, 32 | .31, 29 |
| $b = 1^o$ | 2.07, 30 | 1.68, 31 | 1.23, 31 | .79, 32 | .60, 38 |
|  | 2.17, 30 | 1.94, 31 | 1.36, 31 | .71, 33 | .28, 30 |
| $b = 2^o$ | 1.83, 31 | 1.60, 32 | 1.20, 33 | .83, 35 | .56, 39 |
|  | 1.99, 31 | 1.84, 32 | 1.31, 33 | .70, 36 | .27, 32 |
| $b = 3^o$ | 1.58, 31 | 1.39, 33 | 1.10, 34 | .77, 36 | .52, 39 |
|  | 1.62, 31 | 1.50, 34 | 1.12, 35 | .62, 38 | .25, 32 |
| $b = 4^o$ | 1.13, 34 | 1.05, 35 | .89, 36 | .67, 39 | .47, 42 |
|  | 1.13, 34 | 1.01, 35 | .79, 36 | .44, 39 | .19, 37 |
| $b = 5^o$ | .69, 35 | .69, 36 | .62, 39 | .50, 43 | .38, 45 |
|  | .67, 36 | .58, 38 | .45, 40 | .28, 41 | .13, 37 |
| $b = 6^o$ | .38, 38 | .38, 38 | .38, 42 | .34, 44 | .27, 44 |
|  | .35, 38 | .27, 41 | .21, 38 | .13, 40 | .07, 35 |
| $b = 7^o$ | .18, 37 | .19, 41 | .21, 41 | .19, 44 | .17, 46 |
|  | .16, 39 | .13, 38 | .10, 37 | .06, 36 | .02, 39 |
| $b = 8^o$ | .09, 39 | .09, 43 | .09, 44 | .10, 47 | .10, 47 |
|  | .07, 37 | .05, 36 | .03, 35 | .02, 36 | .01, 37 |

Note. — At each latitude $b$, the upper and lower rows are for the positive and negative longitude $l$ fields respectively. The duration for a lens with mass $m_*$ scales as $m_*^{0.5}$.

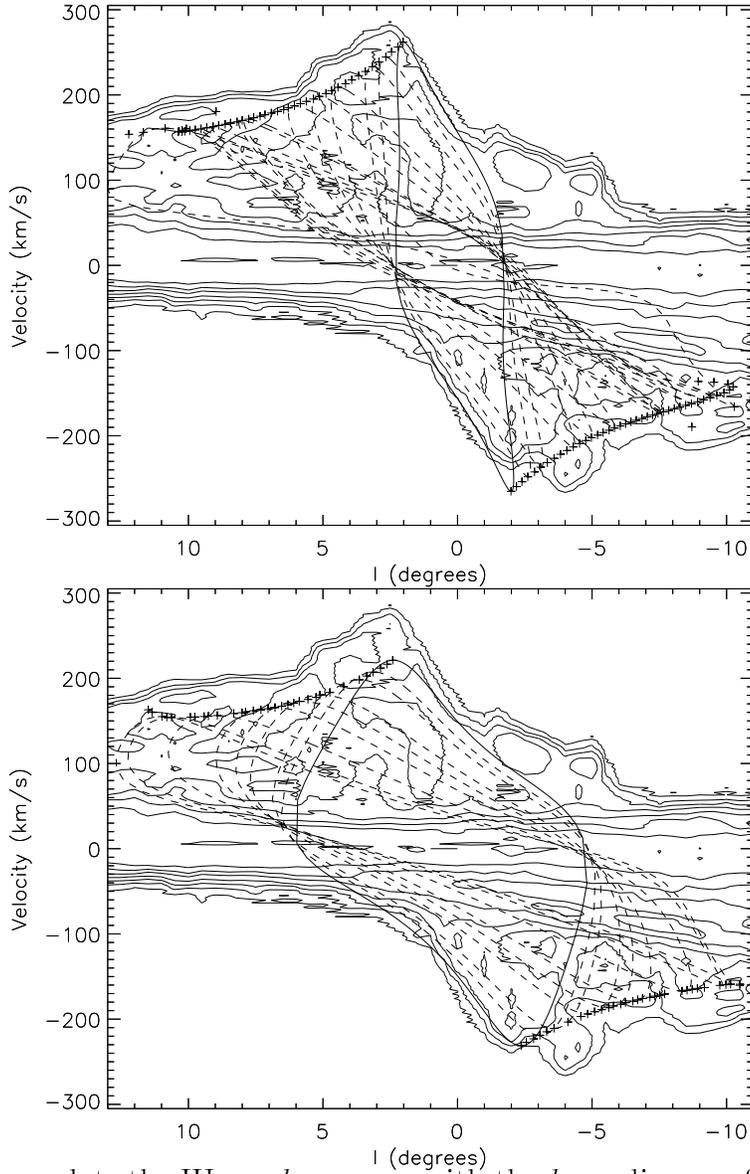

Fig. 1.— overplots the HI gas $l-v$ map with the $l-v$ digram of the $x_1$ orbits for our best model (the upper panel) and the E3 model (the lower panel). The solid line parallelogram-like loop traces the nearly cusped $x_1$ orbits. The crosses mark the positions of terminal velocity of the non-cusped $x_1$ orbits. Note the differences in the shape of the parallelogram and in the fit to the HI profiles. The HI map is based on a FITS file kindly provided by Harvey Liszt.

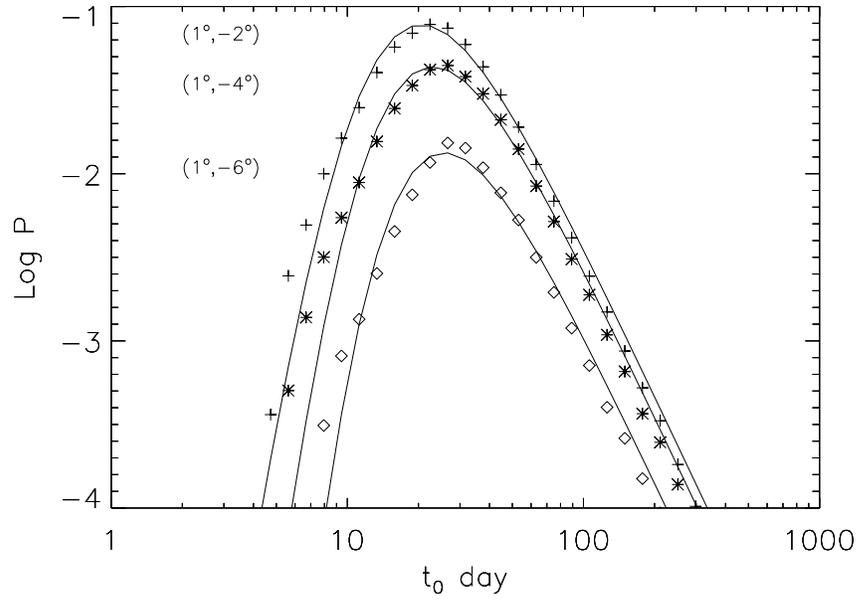

Fig. 2.— compares the parametrized distribution with the rigorously computed event time scale distribution at several bulge fields $(l, b)$. Both assume 100% detection efficiency and $1M_\odot$ bar lenses. The analytical parametrization agrees with rigorous calculation within 20%.

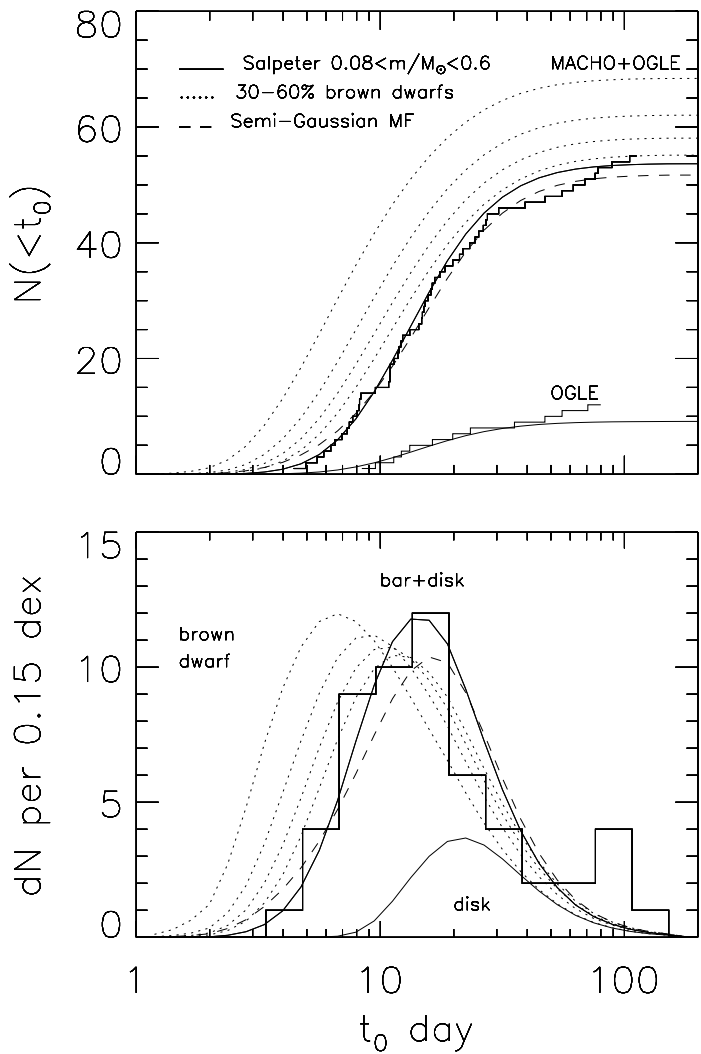

Fig. 3.— plots the cumulative (upper panel) and differential (lower panel) distribution of the time scale for the combined events from MACHO and OGLE (histograms) and our model convolved with detection efficiency and several mass functions. The data are consistent with both a semi-Gaussian mass function with a flat lower end (dashed line) and a Salpeter mass function without brown dwarfs (solid line). Also plotted are predictions of the latter model for the OGLE observation alone (upper panel) and for disk lenses alone (lower panel). Models with 30-60% brown dwarf lenses (the four leftmost dotted lines) are ruled out at $2 - 6\sigma$ levels insensitive to other details of the mass function.